\documentstyle[aasms4]{article}
\def\Ha{H$\alpha$\,}

\def\arcsec{$^{\prime\prime}$}

\def\CaII7291{Ca {\sc II}] $\lambda\lambda$ 7291,7323\ }

\def\OI6300{[O {\sc I}] $\lambda\lambda$ 6300,6364\ }

\def \gta {\mathrel{\vcenter
     {\hbox{$>$}\nointerlineskip\hbox{$\sim$}}}}
\begin{document}

\title
{
Bi-polar Supernova Explosions
}

\author{Lifan Wang\footnote{email: lifan@astro.as.utexas.edu},
        D. Andrew Howell\footnote{email: howell@astro.as.utexas.edu}, 
        Peter H\"oflich,\footnote{email: pah@astro.as.utexas.edu}, 
J. Craig Wheeler\footnote{email: wheel@astro.as.utexas.edu}}

\affil{Department of Astronomy and McDonald Observatory\\
          The University of Texas at Austin\\
          Austin,~TX~78712}

\begin{abstract}

We discuss the optical spectropolarimetry of several core-collapse
supernovae, SN 1996cb (Type IIB), SN 1997X (Type Ic), and SN 1998S (Type IIn).
The data show polarization evolution
of several spectral features at levels from 0.5\% to 
above 4\%. The observed line polarization is intrinsic
to the supernovae and not of interstellar origin. These data suggest 
that the the distribution of ejected matter is highly aspherical. 
In the case of SN~1998S, the minimum major to minor axis ratio must be larger 
than 2.5 to 1 if the polarization is 3\% from an 
oblate spheroidal ejecta seen edge on. A well-defined symmetry axis can 
be deduced from spectropolarimetry for the peculiar Type IIn 
supernova SN 1998S but the Type IIB events SN 1993J and SN 1996cb seem to 
possess much more complicated geometries with polarization position 
angles showing larger irregular variations across spectral features; the 
latter may be associated with large scale clumpiness of the ejecta. The 
observed degree of polarization of the Type Ic SN 1997X is above 5\%. The data 
reveal a trend that the degree of polarization increases 
with decreasing envelope mass and with the depth within the ejecta.
We speculate that Type IIB, Type Ib, and Type Ic may be very
similar events viewed from different aspect angles.

The high axial ratio of the ejecta is difficult to explain in terms of the
conventional neutrino driven core-collapse models for Type II
explosions. Highly asymmetric explosion mechanisms such as the formation
of bipolar jets during core-collapse may be a necessary ingredient for
models of all core-collapse supernovae.

\end{abstract}


\section{Introduction}

Polarimetry probes directly the asphericity of
supernova explosions. It is therefore important to the understanding 
of the supernova explosion mechanism, the pre-explosion environment, and
the interaction of the ejecta with their environment.

Only a few supernovae have been observed with polarimetry.  
SN 1987A represented a breakthrough by
providing the first detailed record of the polarimetric evolution
(Cropper et al. 1988; M\'endez et al. 1988; Jeffery 1991; H\"oflich 1991; 
Wang \& Wheeler 1996).
SN 1993J also provided a wealth of data (Trammell, Hines, and Wheeler; 
Doroshenko, Efimov, \& Shakhovskoi 1995; Tran et al 1996). More recently we 
have begun a program to obtain routine spectropolarimetry on all accessible
SNe. We have used mostly the 2.1 and 2.7 meter telescopes at McDonald 
Observatory. Wang et al (1996) compared broad band polarimetry data 
obtained at the McDonald Observatory with supernova polarimetry data 
published before 1996 and found that all Type II in their sample are 
polarized at about the 1 percent level and that Type Ia are much less 
polarized, less than 0.2-0.3\%. Two Type Ia, SN 1996X and SN 1999by, are 
polarized at levels around 0.2\% near optical maximum. Spectropolarimetry 
of these events
revealed numerous features much narrower than typical P-Cygni lines in the
flux spectra (Wang, Wheeler, \& H\"oflich 1997; Howell et al 2000). 
More data have been acquired since then 
that confirm that core-collapse events are significantly polarized. The 
degree of polarization seems to increase with decreasing masses of the stellar 
envelope, with the highest polarization observed for Type IIB and Type Ib/c events 
(Wang, Wheeler, H\"oflich 1999).

In this {\it Letter}, we discuss the spectropolarimetry 
of several core-collapse supernovae obtained at McDonald Observatory 
(\S 2). We focus on SN 1996cb (\S 2.1), SN 1998S (\S 2.2), 
and SN 1997X (\S 2.3), so far the best observed supernovae in our sample, 
and compare these data with other Type II (\S 2.4). 
The entire polarization data set will be published separately.
Here, we discuss the major features which are most relevant to the 
geometry of the supernova ejecta and to the implications for the 
core-collapse mechanism. A brief summary is given in \S 3. 

\section{Observations}

All of the data on SN 1996cb, SN 1997X, and SN 1998S were obtained at the
McDonald Observatory using the 2.7 meter telescope or the 2.1 meter
telescope. The instrumental polarization of the polarimeter is less than
0.1\%, much lower than the polarization of the supernovae. Due to the
limited light collecting power of our telescopes, we have observed a given
supernova with the same configuration for 2-4 nights to get enough photons
to reduce statistical noise. 
Polarized and unpolarized standard stars were observed in all of our 
observing runs to check the instrumental stability and to correct the 
position angles of the polarimeter. Multiple
sets of data were generally taken with each individual exposure normally
20 to 30 minutes at each waveplate position angle. This is important in our 
spectropolarimetry since the signals are normally
buried in strong noise and independent sets of 
data are useful in removing spurious features.   This procedure also helps to identify
features due to cosmic rays which are extremely numerous at the 2.1 meter
site. The data for each observing run were independently
reduced and then combined to form the final spectra with increased signal to 
noise ratio. The polarization reduction follows the methods
given by Miller, Robinson, \& Goodrich (1988). 
We have also observed a couple of flux standard
stars each observing run to allow for flux calibrations of the supernova
spectra. Details of the observations will be given in a forthcoming paper
(Wang et al, in preparation).

\subsection{SN 1996cb}

SN 1996cb was discovered pre-maximum on 1996 Dec. 15 (Nakono 1996). 
SN 1996cb in NGC 3510 is a Type IIB supernova similar to SN 1993J.
SNe IIB have lost most of their hydrogen envelopes and have only
a tenuous hydrogen layer surrounding them before the explosion. Early 
spectroscopic data are discussed in a recent paper by Qiu et al. (1999). 
The supernova reached optical maximum around 1996 Jan. 2. Despite
some minor differences pointed out by Qiu et al. (1999), the spectroscopic
and photometric behavior of SN 1996cb and SN 1993J are remarkably similar.
Such similarity is also shared by another Type IIB SN 1996B. Both SN 1996cb 
and SN 1993J showed evidence of clumpiness in the ejecta (Wang \& Hu 1994; 
Spyromilio 1994; Qiu et al. 1999), and both SNe were positively detected in 
the radio (Van Dyk et al 1996). Details of SN 1993J can be found in Wheeler \&
Filippenko (1996).
 
Polarimetry data of SN 1993J were obtained around optical maximum.
Trammell et al. (1993) reported data taken  at the McDonald 
Observatory on 1993 April 20. These data were analyzed in detail 
by H\"oflich et al. (1996) where asymmetric helium cores were assumed to 
produce the observations. More spectropolarimetry of SN 1993J was 
reported in Tran et al. (1997). The degree of continuum polarization is an 
evolving function of time and is around 1.7-2\% around 1993 April 20 - 
suggesting a highly distorted envelope with major to minor axis ratio 
$\gta 1.4$ (H\"oflich, 1991). The supernova must be 
viewed nearly edge on for an oblate spheroid or pole on for an
prolate spheroid according to models of H\"oflich et al. (1996). In contrast 
to the well defined trace on the Q-U plot observed for SN 1998S 
(cf \S 2.2, Fig. 3), the position angles of both SN 1993J and SN 1996cb show 
large irregular variations across 
spectral lines. This is consistent with the clumpy ejecta revealed by
spectroscopic observations. These must be large scale
chemical or ionization clumps. Large numbers of small clumps are unlikely to 
be efficient polarizer.

The real surprise is that the two supernovae are not only spectroscopically 
similar but also spectropolarimetrically similar. 
Figure 1 compares the polarimetry of SN 1996cb obtained on 1996 January 5 
with that of SN 1993J taken on 1993 April 26. The SN 1993J data were obtained
at the Kitt Peak National Observatory. Sharp polarization changes were seen 
across the \Ha\ and He 5876 lines. The variation of the polarization
is as large as 1.5\%  in both supernovae. 
In addition, data taken after optical maxima show that 
the polarization of both SN 1993J and SN 1996cb tends to grow with time
suggesting that the asymmetries are larger deeper within the ejecta. 

The fact that SN 1993J and SN 1996cb show similar spectropolarimetry suggests
strongly that the two supernovae shares a similar geometry with a similar
orientation to the observer. The nearly identical polarization spectra 
argues also that, contrary to what has been derived by assuming the 
emission component of the H$\alpha$ line is completely de-polarized by the 
stellar ejecta (Trammell et al 1993; Tran et al 1997), the interstellar 
polarization is, in fact, negligible for both supernovae. It is worth noting 
that the pre-maximum spectroscopic data of another Type IIB, SN 1996B, taken
at McDonald Observatory again bears striking similarities to those of SN 1993J 
and SN 1996cb. 

A natural question to ask is why these supernovae are so similar despite the
fact that they are so highly aspherical and clumpy? The complementary question is 
whether these supernovae are still Type IIB when viewed at different viewing angles. 
Our guess is that
the fact that they are so similar means that they are likely a select 
sub-group of supernovae of a more common phenomena. One possibility is 
that SN Ic, SN Ib, and SN IIB are actually very similar events viewed at 
different angles. The ejecta may have a tenuous helium-enriched hydrogen 
disk which is difficult to observe in the polar directions and can only easily be seen 
close to edge on. The thickness of the hydrogen disk would be related to the 
relative rate of occurrence of SN Ic, Ib, and IIB. More data, especially 
spectropolarimetry, will help to create a unified picture for the SN IIB, and 
perhaps also of SN Ic, and Ib.

At late times, SN 1993J evolved through a remarkable path by showing strong
broad hydrogen Balmer emission lines. If the above picture is correct, then 
we must ask why such broad hydrogen Balmer emission lines have not been detected 
from other Type Ib/c, or from SN 1996cb. Perhaps SN 1993J had 
a relatively massive circumstellar wind before explosion.

\subsection{SN 1998S}

SN 1998S in NGC 3877 was discovered on 1998 March 2.68 by the Beijing 
Astronomical Supernova Search program (Li, Li, \& Wan 1998). It 
was classified as a Type IIn supernova by Filippenko \& Moran (1998) with 
strong emission lines and no obvious absorption features. It reached 
optical maximum around March 20. The optical spectra of SN 1998S showed 
dramatic evolution during the first 3 months. A few days after discovery, 
the spectra consisted of a blue continuum superimposed with many broad 
emission lines with no apparent absorption lines. The lines can be 
identified with emission from H, He, and \ion{C}{3}. The early time 
spectra are remarkably similar to that of a Wolf-Rayet star 
(Garnavich, Jha, \& Kirshner 1998). The spectral profiles of all of the 
emission lines are very similar although they arise from ions with 
different ionization levels. 
Leonard et al. (1999) discussed high quality spectropolarimetry of SN 1998S
obtained at the Keck II 10-meter telescope on 1998 March 7 UT about two 
weeks prior to maximum light.  They found that
the polarization spectrum is characterized by a flat continuum with distinct
changes in polarization associated with both a broad 
(FWZI $\sim$ 20,000 km/sec) and a narrow (FWHM $<$ 300 km/sec). 

\subsubsection{Spectropolarimetry Data at the McDonald Observatory}

Our spectropolarimetry were obtained using the 2.1 meter telescope of the 
McDonald Observatory with the Imaging Grism Polarimeter (IGP) on two 
observing runs each lasting 4 nights. Two sets of data were acquired 
in the first run, both on the night of 1998 March 30, about 10 days after
optical maximum.  Each set of data 
consists of four exposures with exposure time of 20 minutes. Four independent
sets of data were obtained during the second run from 1998 April 29 to 1998 
May 2. Each set of data again consists of four 20 minutes exposures. All of the
data were taken with a slit width of 2\arcsec\ and have a spectral 
resolution of about 12\AA. 

The March 30 and May 1 data were taken at about 10 and 41 days after optical 
maximum of the supernova, respectively. The latter data were taken just
before the light curve began to decline rapidly from the plateau
(Garnavich, private communication 1999).  The supernova showed dramatic 
evolution during this period as is shown in Figure 2 where both the spectral 
and polarimetric behavior had changed dramatically. The March 30 data show an
abnormally strong Si II 6355 \AA\ line and a weak \Ha\ line on top of 
a blue continuum. The velocity of the absorption troughs of the 
P-Cygni features is typically 3200 km $s^{-1}$. The emergence of the strong
lines from Fe II, Sc II, Si II indicates that the photosphere has receded
into slower moving envelope material. The presence of narrow P-Cygni features
indicates that there is slow moving matter, perhaps of circumstellar origin, 
outside the rapidly expanding ejecta.

Understanding spectropolarimetry may be critical in unraveling the
complicated nature of this supernova. As shown in Figure 2. 
A strong de-polarization is observed across the \Ha\ line, meaning
a substantial fraction of the polarization is associated with 
scattering through material in the vicinity of the supernova rather than 
merely by interstellar dust particles on the line of sight to the supernova or 
the scattered light by interstellar dust particles that are very far away 
from the supernova. 

One group of models assumes that the polarization arises from a combination of 
Thomson and line scattering through the aspherical supernova envelope. 
This mechanism has been thoroughly explored in several recent studies 
(H\"oflich 1995; H\"oflich et al 1996) and seems to be successful in 
modeling the spectropolarimetry of SN 1993J. Complete de-polarization by 
line scattering is assumed in these models. The continua are polarized 
by electron scattering whereas the de-polarization across spectral lines 
due to line scattering creates spectral structures across strong P-Cygni 
lines such as \Ha. Another group of models assumes that the supernovae 
are associated with an aspherical dusty circumstellar environment 
which is capable of scattering a significant fraction of the photons from 
the supernovae (Wang \& Wheeler 1996). Because the dust particles are 
located at different 
distances from the center of the supernova, the observer sees photons 
scattered along different light paths at a specific epoch. A rapidly evolving 
supernova spectrum will thus produce a rapidly evolving polarimetry spectrum. 
Note that the degree of polarization is mostly 
determined by the evolution of the supernovae and to a lesser degree 
on the scattering properties of the dust and thus it does not necessarily 
produce higher polarization in the blue than in the red as the conventional
perception would suggest. This second mechanism was applied to SN 1987A 
(Wang \& Wheeler 1996) and 
produced reasonable fits to the observations. Due to the lack of
observational data, it is hard to determine which is a better mechanism in
most cases. In many cases it may be a combination of scattering through both
the supernova ejecta and the circumstellar dust.

The strong de-polarization across 
the \Ha\ line in the March 30 spectrum of SN~1998S indicates clearly that resonance 
scattering through the hydrogen rich envelope is important. Indeed, the
pre-maximum spectropolarimetry reported in Leonard et al. (1999) shows a flat
polarization spectrum for the continuum which is unlikely to be reproduceable
in the circumstellar dust model but is exactly what one would expect if the 
polarization arises from electron scattering within the supernova ejecta.

In terms of the electron scattering model, the March 30 data would 
be understood in a simple model in which the supernova ejecta and hence the 
photosphere is highly aspherical, similar to the models that were
applied to other Type II supernova such as SN 1987A and SN 1993J. 
Leonard et al. (1999) found the overall 
level of polarization to be about 2\% two weeks before maximum. Both our March 30 and May 1 data 
are taken post optical maximum and the supernova has evolved significantly.
Notably, numerous broad absorption lines are detected in our data and 
polarization spectra have revealed more features than the Keck II data.
Our data are unfortunately unable to resolve the narrow (FWHM $<$ 300 km/sec)
\Ha\ line detected by the Keck II telescope.

Some important features revealed by the spectropolarimetry can be summarized
as follows: 

(a) The degree of polarization evolves with time. Polarization variations
are observed across strong emission/absorption lines such as H$\alpha$,
Si II, He I, Fe I, and Fe II. Unlike the pre-maximum data 
(Leonard et al. 1999), the continuum polarization is no longer flat in our 
post-maximum data. The degree of polarization is the highest in the blue 
($\sim 1.6\%$) and is the lowest at around
5500 \AA\ ($\sim 0.4\%$); (b) The position angles of the polarization vector
evolved significantly during the two epochs of our observations compared to 
the earlier Keck II data; (c) A sharp change of polarization
position angle by 90 degrees is observed across the H$\alpha$ line in the May 1
spectrum; (d) the polarization position angle changed by nearly 90 degrees from
red to blue in the May 1 spectroscopic data.

\subsubsection{How Large is the Intrinsic Polarization?}

It is important to separate the polarization component due to interstellar
dust along the line of sight to the supernova from the components which
are intrinsic to the supernova. This is always difficult and depends on 
certain assumptions of the polarization properties of the emission/absportion
lines and the continua. Leonard et al. (1999) were able to separate the 
H$\alpha$ lines into a narrow (unresolved) and broad component which show
different polarizations. The authors then tried to derive the interstellar
polarization by assuming either the narrow or the broad component is 
intrinsically unpolarized and the observed polarization is due entirely to
interstellar polarization.  Lacking a proper knowledge of the 
origin of the narrow and broad components, it is hard to justify which 
assumption is better. 

The maximum degree of interstellar polarization is found to be 9 $\times$
E(B-V) from observations of interstellar polarization in the Galaxy 
(Serkowski et al 1975). The
maximum achievable polarization corresponds to the configuration in which
all the dust particles are aligned to the optimum angle for polarization.
The circle in Figure 3 shows the limit of allowable interstellar polarization
according to the Serkowski relation. It is, however, uncertain because of the
uncertainty of the derived extinction. 

Certain assumptions must be made in order to estimate the interstellar 
polarization. We will assess the interstellar polarization via a 
different approach by relaxing the constraints on the polarization properties 
of emission lines and assume only that at each specific epoch, the 
polarization produced by the supernova ejecta has a single polarization 
position angle independent of wavelength. This can be achieved, for instance, 
if the polarization is produced purely by electron/dust scattering in the 
vicinity of the supernova with no large scale chemical inhomogeneities. 

The Q-U plot in Fig. 3 is consistent with our assumptions. The polarization
at different wavelengths falls roughly on a straight line on the Q-U plot which
is typical of the combination of two polarization vectors of which one is
nearly constant (like interstellar polarization) while the other highly 
variable. Under this assumption, the interstellar polarization must lie 
somewhere along the line drawn by the observed Q and U vectors. The allowable
areas are marked in Fig. 3 as A and B. 

We need theoretical input to determine whether A or B is a more 
reasonable area for the interstellar polarization. For the 1999 May 1 data,
the question is equivalent to asking whether the shorter wavelength 
or the longer wavelength is more likely to be highly polarized.
We believe that the shorter wavelength suffers severe de-polarization due to
numerous Fe II, and Sc II lines whereas there are few strong lines at 
wavelengths longer than H$\alpha$ and therefore the longer wavelength side 
is more likely to have higher polarization. This limits the possible 
interstellar component to area A. In the subsequent discussion we will take 
Q $\sim$ 0.7\% and U $\sim$ -1.9\% for the interstellar polarization. Note also
our determination of the interstellar polarization yields a position angle of 
-35 degrees for the interstellar polarization.

This determination of the interstellar polarization leads to intrinsic 
polarization of the supernova of around 1\% for the March data and over 3\%
a month later. To produce such a high degree of polarization for density 
profiles which resemble those of SNe~II, the 
major/minor axis ratios must be larger than 1.2 to 1 and 2.5 to 1   
for the early and late observations, respectively,
if the ejecta distribution is an oblate spheroid viewed edge on.
Any other viewing angle $\Theta $ will require larger axis ratios because
the polarization scales as $ \approx P(equatorial)  sin^2 \Theta $.
Once again, we note that the polarization increases with time suggesting
greater distortions the deeper one goes in the ejecta.
 The degree of polarization is extraordinarily large and clearly shows
a bi-polar structure for the ejecta. 

Another Type IIn, SN 1994Y, was also observed in our program close to optical
maximum and showed similar behavior to SN 1998S (Wang et al. 1996). The 
degree of polarization 
variation across the \Ha\ line is larger than 1.5\% in that case, and 
by using the same method for separating the interstellar polarization, the
intrinsic polarization can be as large as 3\%. This is consistent with 
SN 1998S. These are the only two Type IIn supernovae with polarimetry
observations so far.
We note that, despite their extraordinarily flat light curves, Type IIn 
supernovae are probably an extreme class of objects which have lost most
of their massive envelope and the ejecta mass can be less than 1 M$_\odot$ 
(Chugai, \& Danziger 1994). 
This is consistent with the Wolf Rayet features of SN~1998S and of
its interpretation as a Wolf Rayet star by Gerardy et al. (1999).
 
\subsection{Type Ic SN 1997X}

SN~1997X is one of the best observed supernovae in terms of polarimetry. The 
observed total degree of polarization is as high as 7\%. A significant fraction
of the polarization may be of interstellar origin. The polarized spectra do
show clear spectral features across the strong He I 5876, 6678 lines (Wang,
Wheeler, H\"oflich 1997). Although it is hard to separate the intrinsic and 
interstellar polarization, it is also clear that a large faction of the
of the polarization must be associated with the supernova ejecta. Fig. 4
shows the time evolution of the continuum polarization. A polarization level
of at least 4\% can be attributed to an asymmetric ejecta 
with axis ratios larger than 3 to 1 for oblate spheroids.

\subsection{Other Type II supernova}

We have observed a few normal Type II plateau supernovae which are believed to have
more massive stellar envelopes. The degree of polarization is not as large
as those of SN 1987A (at late times), SN 1993J, SN 1996cb, and SN 1997X and
is typically around 0.5-1.5\%. These supernovae and the detected levels of 
polarization are: SN 1995H, 1\%; SN 1995V, 1.5\%; SN 1996W, 0.7\% which
corresponds to deviations from symmetry between 10 to 30 \%.
We expect relatively low polarization because of the damping effect of
the hydrogen envelope.  Even these moderate levels of polarization for 
SN~IIP suggest a strong asymmetry deep within the ejecta.  So far, we
have observed no exceptions to the statement that all core-collapse
supernovae are substantially polarized and hence substantially asymmetric.
The recent supernova SN 1999em was probably a supernova with a massive 
envelope.
Pre-maximum polarimetry at the Keck II telescope showed no intrinsic 
polarization down to 0.1\% suggesting either a spherical ejecta or a 
nearly pole-on viewing angle (Leonard, Filippenko \& Chornok 1999).
Alternatively, the polarization could also be suppressed if the 
density slope is very steep as indicated by the small differential 
Doppler shifts of the lines of the Balmer series (H\"oflich 1995). 
This null result may thus be due to the early epoch
of observation.  It will be interesting to obtain post-maximum data to
probe deeper into the ejecta to regions more directly 
related to the explosion mechanism.

\section{Discussion}

These polarization observations reveal that core-collapse supernovae
are generally polarized and are highly aspherical. The degree of polarization
evolves with time and generally increases after optical maximum. 
The polarization is 
also anti-correlated with the mass of the remaining hydrogen shell. 
The trend is that highest polarizations are observed for 
supernovae that have lost most of their hydrogen envelope before explosion.
This trend is illustrated by the sequence Type II SN~1987A, 
Type IIn SN~1994Y and SN~1998S, Type IIB SN~1993J and 1996cb, 
and Type Ic SN~1997X. From their relatively large variations of the 
polarization position angles across spectral lines, Type IIB events 
seem to possess not only large global asymmetry but also 
strong large scale chemical or ionization clumps.

Polarization as large as 4\% along a constant orientation axis 
requires a strongly bi-polar departure from
spherical symmetry.  How was such a geometry produced? Does the 
spectropolarimetry reveal an intrinsically aspherical supernova explosion 
explosion? One possibility
might involve massive stellar disks surrounding the supernova so that even
though the explosion is spherical, the ejecta/disk interaction could distort
the photosphere. The increasing degree of polarization after optical maximum
observed in SN 1987A, SN 1993J, SN 1996cb, SN 1997X, and SN 1998S
strongly favors an aspherical explosion. 

Both the temporal evolution and the dependency on ejecta mass of the
polarization are consistent with aspherical explosions. The asphericity
is expected to be the largest near the center of the ejecta which is normally 
revealed only by observations past optical maximum.  The asymmetry is also 
expected to have the largest effect on the ejecta geometry for bare-core 
progenitors such as those for Type Ib/c since a massive hydrogen envelope
tends to smear out the asphericity.

These observations argues strongly {\it against} the conventional picture
for Type II supernovae explosions. It is hard to imagine that neutrino-driven 
explosions can produce such large asymmetries throughout the ejecta. 
We note that the neutrino-driven explosion models have so far failed
to produce robust explosions. 
To produce the bi-polar structure we observe with the polarimetry,
a strong asymmetry must be imposed and maintained for a time substantially
long compared to the dynamical time of the progenitor.  This criterion can 
be satisfied by the production of a sustained jet from deep within the
core.  The recent jet-driven explosion models by Khokhlov et al. (1999) 
and MacFadyen \& Woosley (1999) provide a more promising approach
to accounting for the polarization.  We note that since the polarization
is ubiquitous these jets must routinely arise from neutron stars, not
just in the rare events that produce explosions and black holes.
Asymmetric jets could provide both the bi-polar ejecta and a kick
to the neutron star (Khokhlov et al. 1999).  On the contrary,
asymmetric neutrino emission could deliver a kick to a pulsar, but
it is unlikely to generate the bulk bi-polar nature of the
ejecta that we deduce.

We thank Alex Filippenko and Doug Leonard for useful discussions and 
for providing data before publication. This work is supported in part by 
NSF Grant 9818960, by a grant from the Texas Advanced Research Program
and by NASA through grants LSTA-98-02 and HF-01085.01-96A from 
the Space Telescope Science Institute which is operated by the Association of
Universities for Research in Astronomy, Inc., under NASA
contract NAS 5-26555.

\pagebreak

\noindent
{\bf Figure Captions}\\[1cm]

{\bf Fig. 1...} The total flux and polarization spectra of SN 1996cb are compared to
those of  SN 1993J. 

{\bf Fig. 2...} Spectropolarimetry of SN 1998S.  The left column gives the
total flux spectrum, the percent polarization and the polarization angle
for the data of 1998 March 30, respectively and the right column gives
the corresponding data for 1998 May 1.

{\bf Fig. 3...} The Spectropolarimetry of SN 1998S on the Q-U plane
for 1998 March 30 (left panel) and 1998 May 1 (right panel). 
The outer circles are the limit of the interstellar polarization from
the extinction.  The premaximum data from Leonard et al. (1999) is
labeled by Keck II data.  The circles labeled A and B represent the two
choices for the ISM, of which we prefer A (see text). The 
small colored circles in the May 1 data are from 6397\AA\ to 6802\AA\
in the vicinity of the \Ha\ line. 

{\bf Fig. 4...} The evolution of the continuum polarization (Q, U, and
total polarization P), for SN 1997X 
obtained by averaging from 5000\AA\ to 6000\AA.
Note the large change with time which must be intrinsic to the
supernova and suggests a polarization of order 4\%.

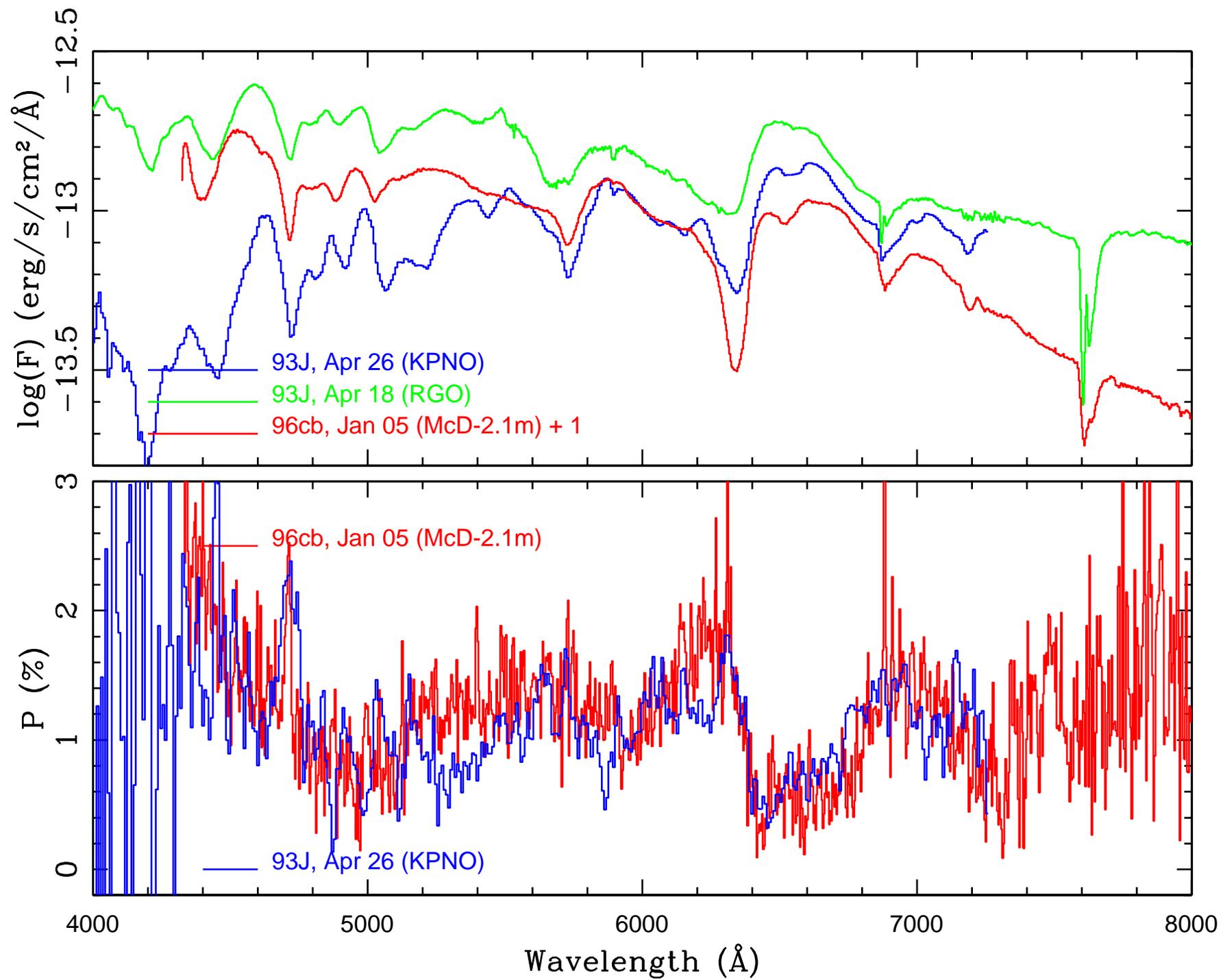
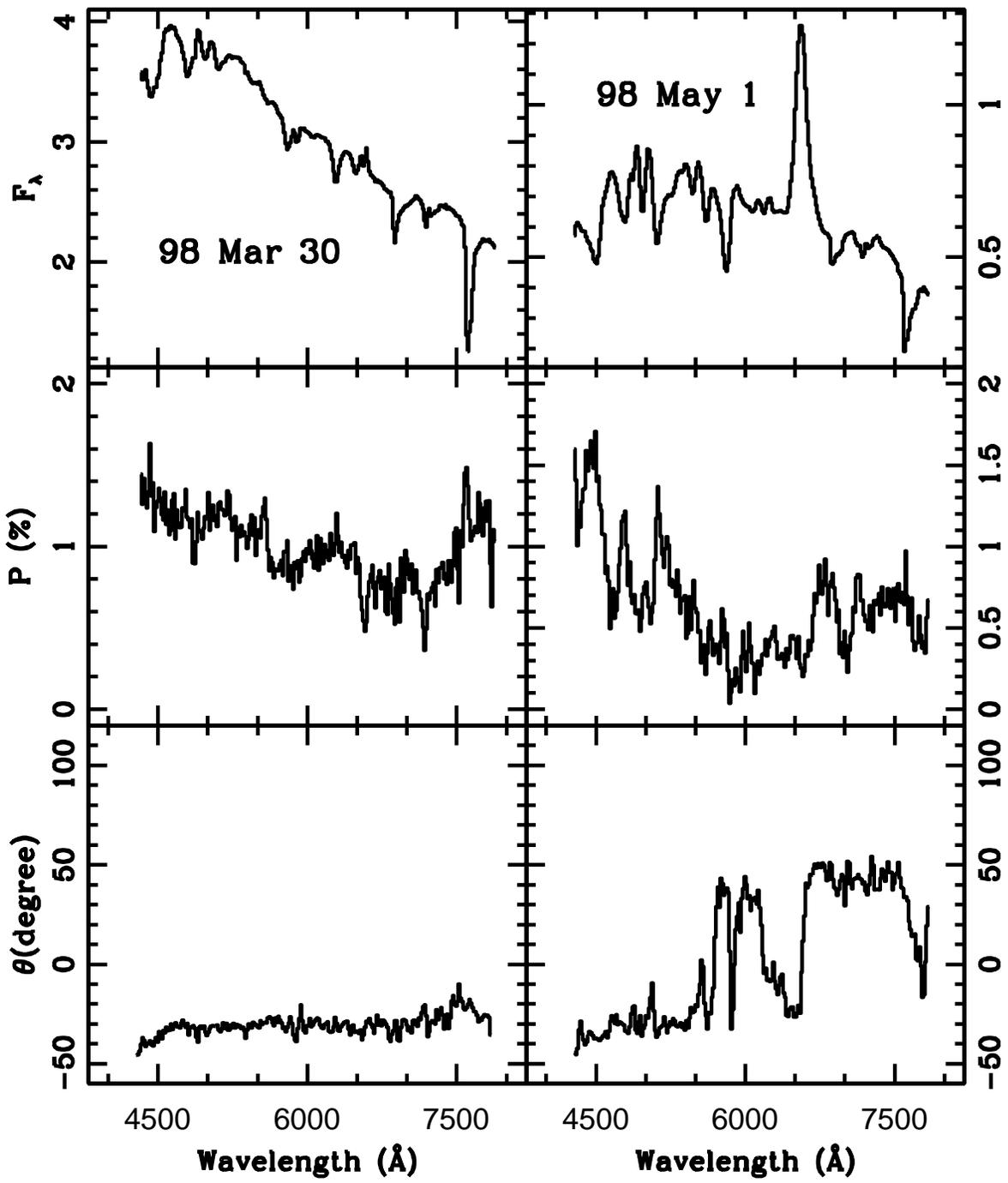
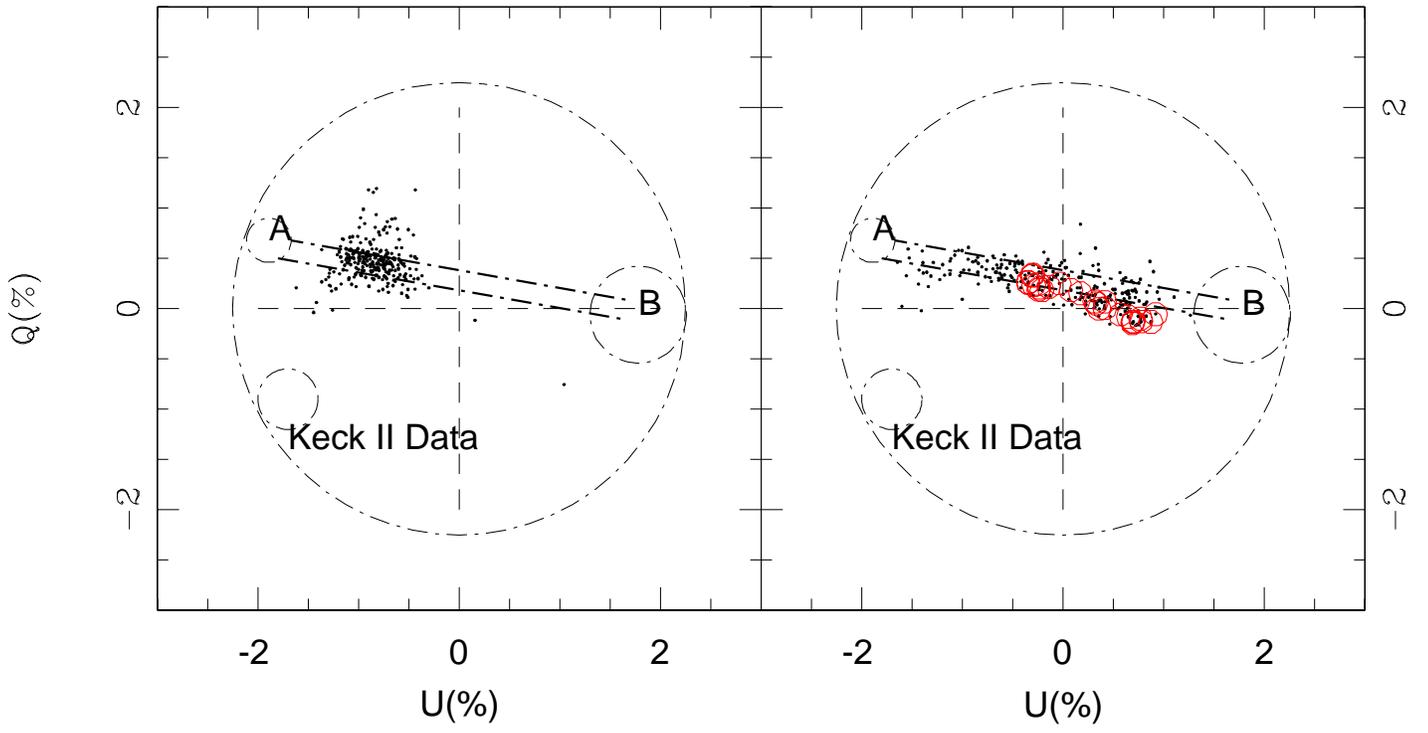
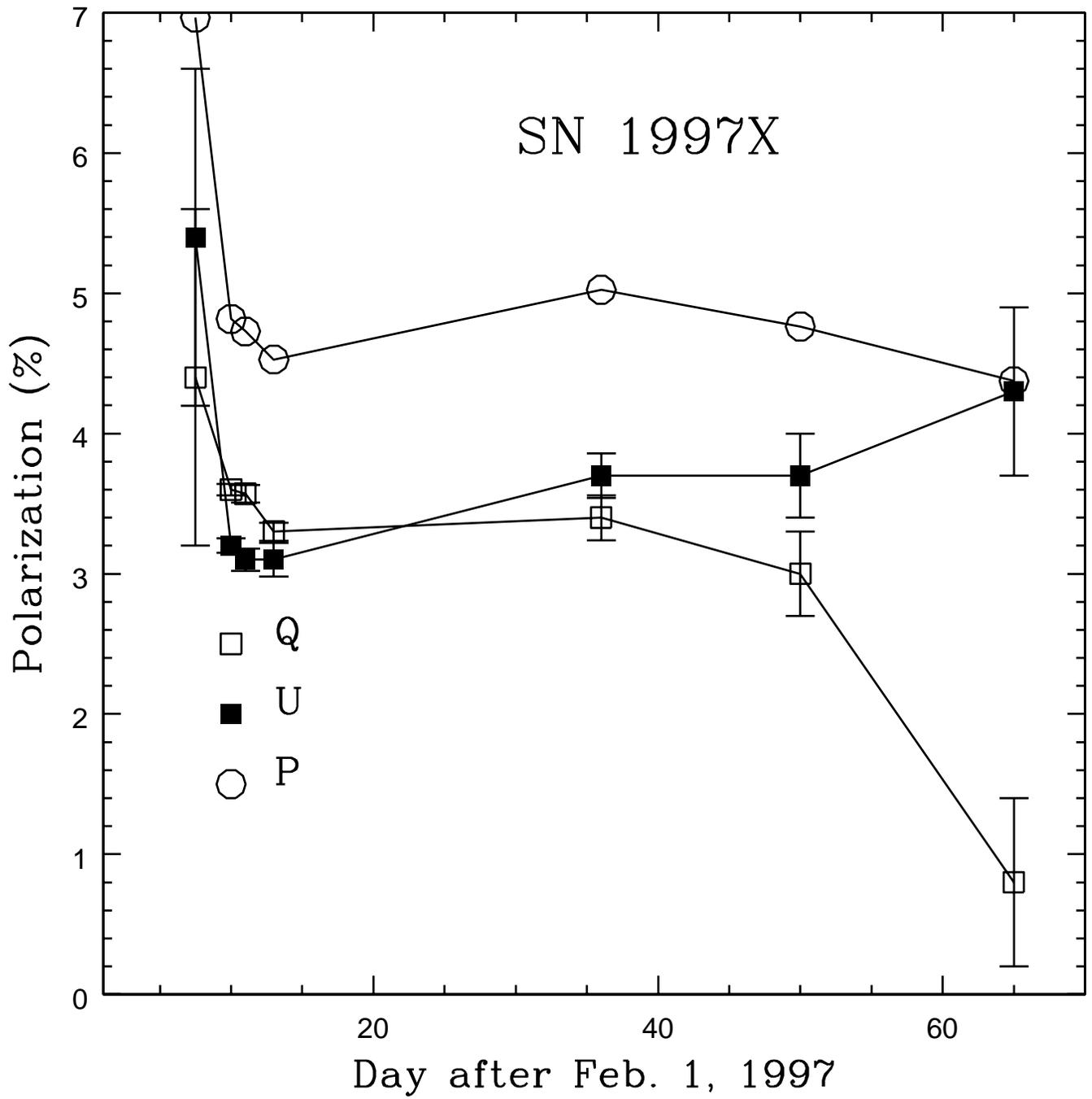
\end{document}